\documentclass[12pt]{iopart}

\usepackage{iopams}
\usepackage{graphicx}
\usepackage{color}
\usepackage{textcomp}
\usepackage{subfig}
\usepackage[pdfusetitle]{hyperref}
\newcommand{\micron}{\textmu{}m}

\hypersetup{
  pdfauthor={Tobin T. Fricke, Nicolas D. Smith-Lefebvre, Richard Abbott, Rana Adhikari, Katherine L Dooley, Matthew Evans, Peter Fritschel, Valery V. Frolov, Keita Kawabe, Jeffrey S. Kissel, Bram J. J. Slagmolen, Sam J. Waldman}
}

\begin{document}
\title{DC readout experiment in Enhanced LIGO}

\author{
  Tobin T Fricke,$^{1,*}$ 
  Nicol\'as D Smith-Lefebvre,$^{2}$ 
  Richard Abbott,$^{3}$
  Rana Adhikari,$^{3}$ 
  Katherine L Dooley,$^{4}$ 
  Matthew Evans,$^{2}$ 
  Peter Fritschel,$^{2}$ 
  Valery V Frolov,$^{5}$ 
  Keita Kawabe,$^{6}$ 
  Jeffrey S Kissel,$^{2}$ 
  Bram J J Slagmolen,$^{7}$ and
  Sam J Waldman$^{2}$
}

\address{
$^1$ Department of Physics and Astronomy, Louisiana State Univ.,
     Baton Rouge, LA %70803-4001 
}
\address {
$^2$ LIGO Laboratory, Massachusetts Institute of Technology,
      Cambridge, MA %02139 
}
\address {
$^3$ LIGO Laboratory, California Institute of Technology,
      MS 100-36, Pasadena, CA %91125
}
\address {
$^4$ Department of Physics, University of Florida,
      Gainesville, FL %32611-8440
}
\address {
$^5$ LIGO Livingston Observatory,
     PO Box 940, Livingston, LA %70754-0940
}
\address {
$^6$ LIGO Hanford Observatory,
     PO Box 159, Richland, WA %99352-0159 
}
\address {
$^7$ Australian National University, 
     Canberra, ACT 0200, Australia
}

\ead{tobin.fricke@aei.mpg.de}

\begin{abstract}
The two 4 km long gravitational wave detectors operated by the Laser
Interferometer Gravitational-wave Observatory (LIGO) were modified in
2008 to read out the gravitational wave channel using the DC readout
form of homodyne detection and to include an optical filter cavity at
the output of the detector. As part of the upgrade to Enhanced LIGO,
these modifications replaced the radio-frequency (RF) heterodyne
system used previously. We describe the motivations for and the
implementation of DC readout and the output mode cleaner in Enhanced
LIGO. We present characterizations of the system, including
measurements and models of the couplings of the noises from the laser
source to the gravitational wave readout channel. We show that noise
couplings using DC readout are improved over those for RF readout, and
we find that the achieved shot-noise-limited sensitivity is consistent
with modeled results.
\end{abstract}


\vspace{1in}
To be published in \CQG

%\vspace{0.5in}
%\texttt{\color{red}{Compiled \today}}

\maketitle 

\section{Introduction}
The Laser Interferometer Gravitational-wave
Observatory\cite{S5InstrumentPaper} (LIGO) operates gravitational wave
detectors at two sites in the United States (at Hanford, Washington
and Livingston, Louisiana) as part of a global network of
observatories with the goal of making a first direct detection of
gravitational waves (GWs) and beginning an era of gravitational wave
astronomy.

The LIGO facilities were designed to accommodate an initial detector
design using the technology available at the time followed by a major
upgrade known as Advanced LIGO\cite{Abramovici1992LIGO}.  The initial
 detectors reached design sensitivity in 2005 and
subsequently completed a two year observational run (S5).  Rather than
continue to run in this configuration until the beginning of the
Advanced LIGO upgrades in 2010, it was decided to upgrade
the detectors opportunistically to an intermediate configuration known as
Enhanced LIGO\cite{JoshSmithEnhancedAdvanced, T050252,
  Adhikari2006Enhanced}.  Enhanced LIGO culminated with LIGO's sixth
science run (S6), which took place between July 2009 and October 2010 using this new
configuration.

This paper describes two of the changes made in Enhanced LIGO:
implementation of an optical filter cavity (the \emph{output mode
  cleaner}) at the interferometer's output port, and implementation of
a new technique to extract the gravitational wave signal from the
interferometer, called \emph{DC readout}.

\section{Experimental arrangement}
\begin{figure}
  \centerline{
    \includegraphics[width=\columnwidth]{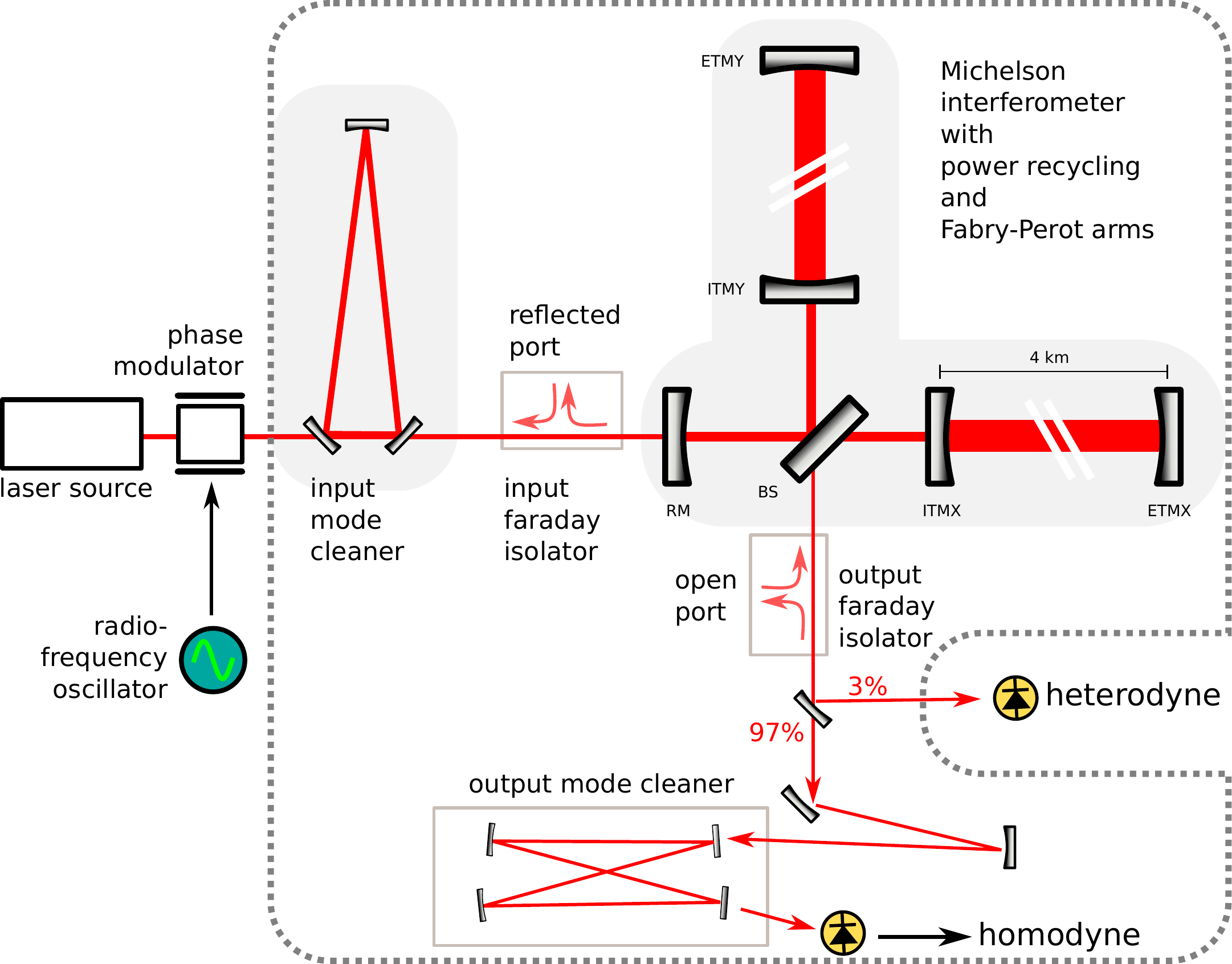}
  }
  \caption{
    \label{fig:dc-readout-diagram}
    Schematic diagram of the main interferometer layout (not to
    scale).  The dotted line represents the vacuum envelope.  }
\end{figure}

Each LIGO detector is a Michelson interferometer with 4 km long arms
consisting of Fabry-Perot resonant cavities (see
figure~\ref{fig:dc-readout-diagram}).  
To increase the amount of laser power at the beamsplitter and in the
interferometer arms, a \emph{power recycling mirror} is used to direct
the light reflected from the Fabry-Perot Michelson back into the
interferometer.  The combination of the power recycling mirror and the
Michelson forms the \emph{power recycling cavity} (PRC).
The positions of the mirrors are combined into several degrees of
freedom: the difference of the arm cavity lengths (DARM), the common
(mean) arm cavity length (CARM), the short Michelson (the difference
of the distances from the beamsplitter to two arm cavity input
mirrors; MICH), and the power recycling cavity (formed by the mean
distance to the two arm input mirrors; PRC).

A suitably polarized gravitational wave incident on the detector
produces differential phase modulation in the two arms, which is
resonantly enhanced by the Fabry-Perot cavities.  The resulting phase
modulation sidebands interfere constructively at the beamsplitter and
exit at the output port, while transmission of the carrier to the
output port is suppressed by operating the Michelson near its dark
fringe.  

To sense the signal sidebands induced by a GW with a photodiode, it is
necessary to introduce an additional local oscillator (LO) field to
produce power variations that are detected by the photodiode.
The initial generation of laser interferometer gravitational wave
detectors used a heterodyne detection
scheme\cite{Fritschel2001Readout} inspired by the Pound-Drever-Hall
technique\cite{Drever1983Laser}.  In this scheme, the local oscillator
field consists of two strong radio-frequency (RF) sidebands separated
from the laser carrier by tens of MHz (25 MHz in initial/Enhanced
LIGO). The interference between these local oscillator sidebands and
the GW-induced sidebands produces a power modulation at the RF
frequency whose amplitude is modulated by the GW amplitude.  The GW
signal is recovered by electronically demodulating the photodiode
signal at the RF frequency.  This scheme is known as \emph{RF
  readout}.
Typically (and in LIGO), the RF sidebands are created by phase
modulating the laser light before it enters the interferometer.  To
allow these sidebands to reach the output port while simultaneously
suppressing transmission of the carrier, the short Michelson is built
with a macroscopic difference in its arm lengths known as the Schnupp
asymmetry (in initial/Enhanced LIGO, 355 mm).  By detecting light from other
interferometer ports, RF readout can also be used to sense other
degrees of freedom, such as the short Michelson, the length of the
power recycling cavity, and the mean length of the two arm
cavities\cite{Regehr1994Signal}.

An alternative method of recovering the GW signal is to provide a
local oscillator at the carrier frequency, an arrangement known as
homodyne detection.  The interference between the carrier-mode LO and
the GW-induced sidebands produces a power variation on the photodiode
that directly reproduces the GW signal.  \emph{DC readout} is a
special form of homodyne detection where the carrier-mode local oscillator is
produced by introducing a static offset in either the short Michelson
or in the difference between the arm cavity lengths, either of
which will produce carrier light at the output port (but with different
consequences for noise couplings).
DC readout has two significant advantages over balanced homodyne
detection\cite{McKenzie2007Technical}, in which the local oscillator is provided by an independent
path and combined with the signal beam via a beamsplitter: no
additional optical path is needed to deliver the local oscillator to
the output port, and the local oscillator that is generated benefits
from being filtered by the combined action of the power recycling and
arm cavities.

There are also a few potential drawbacks to DC readout as compared to
balanced homodyne detection.  In the DC readout configuration, the
phase of the local oscillator relative to the signal field cannot be
tuned directly; it is determined by the relative amplitudes of the
intentional offset and the contrast defect.  The static carrier field
at the interferometer output port results in an unintended coupling of
beam motion to the wavefront sensor at the output port, used by the
interferometer's automatic alignment
system\cite{Barsotti2010Alignment}.  Finally, detuning the arm
cavities from resonance introduces an optical spring effect, which may
lead to excessive coupling of laser intensity noise via radiation
pressure effects in some configurations\cite{ChaibiOptomechanical}.

\begin{figure}[t]
\centerline{\includegraphics[width=0.5\columnwidth]{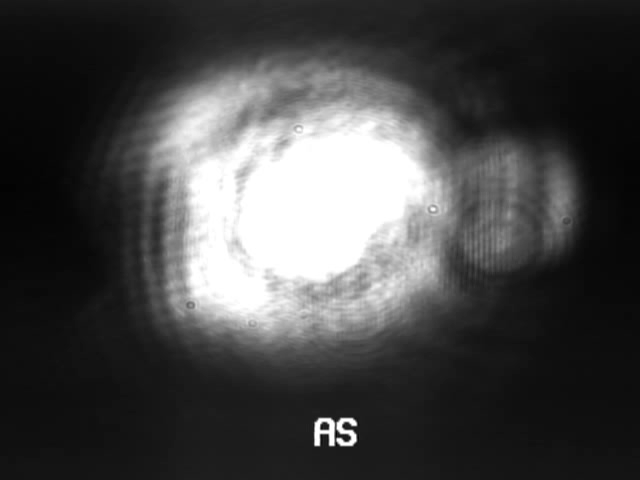}}
\caption{\label{fig:as-spot}Image of the beam spot at the L1 output
  port taken using a CCD camera.  This image is saturated in the
  central portion but emphasizes the spurious higher order modes
  surrounding the fundamental Gaussian, including contributions
  from both the carrier and the 25 MHz sidebands.}
\end{figure}

The optical fields in the interferometer are intended to exist in only
the fundamental Gaussian spatial mode.  A critically coupled resonant
cavity, the input mode cleaner, is used to attenuate any higher order
spatial modes produced by the laser before the field is incident on
the interferometer\cite{Rudiger1981Mode,Dooley2011InputOptics}.  Despite having an essentially pure input beam,
imperfections in the interferometer optics lead to the production of
higher order spatial modes in the interferometer; this effect is
particularly egregious for the RF sidebands in the power recycling
cavity\cite{Gretarsson2007Effects}.  As a result, the output beam
(depicted in figure~\ref{fig:as-spot}) is no longer in the pure
Gaussian mode.  These higher order spatial modes are detrimental as
they generally produce no useful signal, but contribute additional
photon shot noise, increase the power that needs to be detected, and
exacerbate noise couplings.  To mitigate these effects, an
\emph{output mode cleaner} (OMC) was installed at the output port.
This critically-coupled optical filter cavity attenuates higher order
spatial modes before the beam is detected by a pair of photodiodes.
In DC readout, the OMC is also used to remove the RF sidebands, which
are still needed in the interferometer to sense other degrees of
freedom, but would only be detrimental to the DC readout signal.

% Prior DC readout projects
Both DC readout and an OMC have been implemented previously, but this
was the first use of the combination on a km-scale GW detector with
Fabry-Perot arms.  Before implementing these technologies on the main
LIGO interferometers, they were prototyped at the 40 meter prototype
interferometer at Caltech\cite{Ward2008DC,RobWardThesis}.  To see the
full benefit of DC readout required an interferometer with
long arms in order to achieve filtering of the laser carrier at the
frequencies of interest; Enhanced LIGO provided validation of the
low-noise performance of DC readout.

DC readout was simultaneously implemented and is currently used at the
GEO 600 detector\cite{GeoDC,Prijatelj2010,Degallaix2010Commissioning,Prijatelj2012Output}                                                                    
near Hannover, Germany.
The current configuration of the Virgo detector near Pisa, Italy incorporates an output mode cleaner
but uses RF heterodyne readout\cite{Acernese2008Virgo}. Both Advanced
Virgo\cite{AdvVirgoBaselineDesign} and Advanced
LIGO\cite{Abbott2008Advanced} will use DC readout to sense the primary
GW signal.
% Also cite: RF OMC at Hanford
% \cite{Kawabe2004Excess,Betzwieser2004Study}.

\section{Motivation}

The RF readout technique was used successfully in initial LIGO to
achieve the instruments' design sensitivity.  A number of reasons,
outlined below, motivated the switch to DC readout with an OMC:

\subsection{Improved noise couplings} 
The combination of the power recycling cavity and the arm cavities
acts as a single resonant cavity with an effective linewidth of
approximately 1 Hz (the so-called coupled cavity pole) around the
laser carrier.  The RF sidebands, which are not resonant in the arms,
experience no such filtering in the band of interest\cite{Camp2000Analysis}.  DC readout
exploits this filtering by using the carrier light that has circulated
in the interferometer as the local oscillator.  The result is that
coupling of noises on the input beam to the gravitational wave readout
can be greatly reduced.

%% Oscillator noises have been an issue for LIGO in the past
%% \cite{RanaThesis,Ballmer2006LIGO}; this was mitigated by using a low
%% noise crystal oscillator from Wenzel Associates, Inc in place of a
%% general purpose RF function generator. 

\subsection{Spatial overlap}
Because the signal beam (resonant in the arm cavities) and the RF
sideband beam (resonant in the power recycling cavity) are resonant in
different cavities, they will emerge from the interferometer in
different spatial modes if these cavities are not perfectly
matched. This imperfect spatial overlap was a significant problem
during initial LIGO.  For instance, a measurement in 2003 found only
half the expected optical gain\cite{Fritschel2003Shot} (the ratio of
photodiode signal produced to modulation of the differential arm
length).  This was alleviated in part through the use of a thermal
compensation system (TCS), which projected light from a CO$_2$ laser
onto the test mass optics to adjust their effective radii of
curvature, compensating for thermal lensing\cite{Ballmer2006LIGO}.
However, spurious fields still caused problems by producing a large
signal in the uncontrolled quadrature of the heterodyne readout
(AS\_I); left uncontrolled, this signal would saturate the photodiode
electronics.  This was partially mitigated by an electronic servo
which cancelled this signal in the photodiode head.

Both DC readout and an OMC mitigate problems with the spatial overlap
of the LO and signal beams.  In DC readout, the local oscillator and
the signal beams resonate in the same cavities, so spatial overlap
comes naturally.  With either DC or RF readout, an OMC, matched to the
spatial mode of the signal beam, can be used to select the signal beam
and the part of the LO with good spatial overlap.

\subsection{Excess power}
To cope with the excess power due to higher-order modes at the
interferometer output port, initial LIGO split the light at the
detection port onto four detection photodiodes.  Increasing the
interferometer input power (to improve the sensitivity in the
shot-noise-limited frequency band) would require a commensurate increase in
the number of photodiodes at the output port and their associated
electronics.  This is of particular concern for Advanced LIGO, which will
utilize a 200 Watt laser source in place of Enhanced LIGO's 35 Watt laser.
An output mode cleaner attenuates the higher-order modes, reducing the
power that needs to be detected.

\subsection{Homodyne SNR advantage} 
Homodyne detection confers a fundamental improvement in
signal-to-noise ratio compared to RF readout at shot-noise-limited
frequencies by a factor $\sqrt{3/2}$ (for the same power circulating
in the interferometer).  The extra noise in heterodyne detection is a
result of cyclostationary shot noise\cite{Niebauer1991Nonstationary}.

\subsection{Squeezed vacuum injection} 
Squeezed vacuum injection is an attractive means to decrease the
photon quantum noise in future interferometers by manipulating the
quantum state of the vacuum field that enters the interferometer
through the output port.  Squeezed vacuum injection is more feasible
in conjunction with homodyne detection than with RF readout, since it
requires squeezing in only the audio band rather than at both audio
and RF frequencies\cite{GeaBanacloche1987Squeezed,
  Chelkowski2007Coherent}.  Shot noise reduction via squeezed vacuum
injection has been demonstrated at GEO\cite{GeoSqueezingNature2011}
and an effort is currently underway to implement it at LIGO
Hanford\cite{H1SqueezerProposal}.

\section{Implementation of DC readout}

\begin{figure}[t]
  \subfloat[][Schematic of OMC bench]{
  \includegraphics[width=0.5\columnwidth]{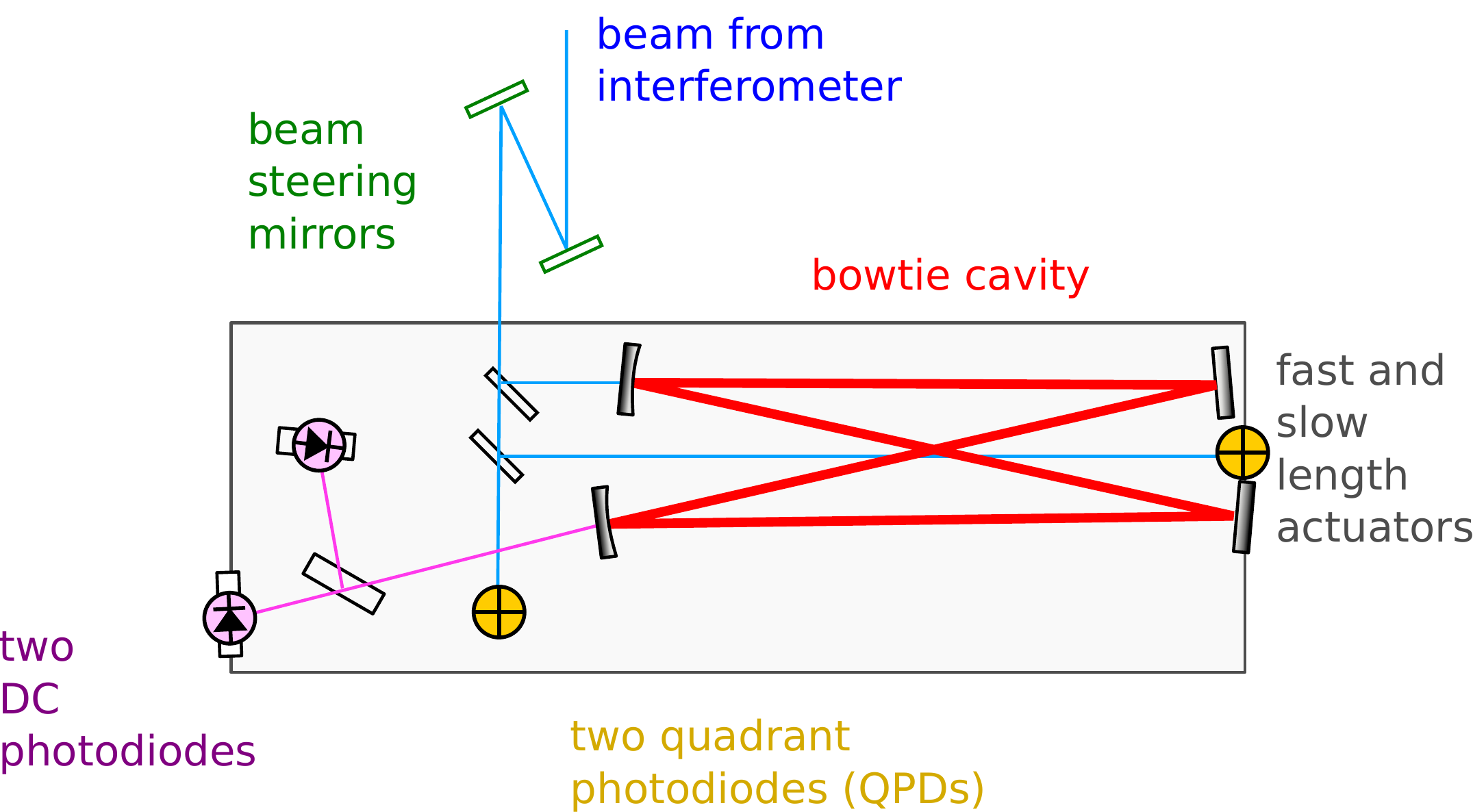}
  \label{fig:omc-diagram}
}
\subfloat[][\label{fig:OMC-chamber}Photograph of installed OMC and suspension]{
  \includegraphics[width=0.5\columnwidth]{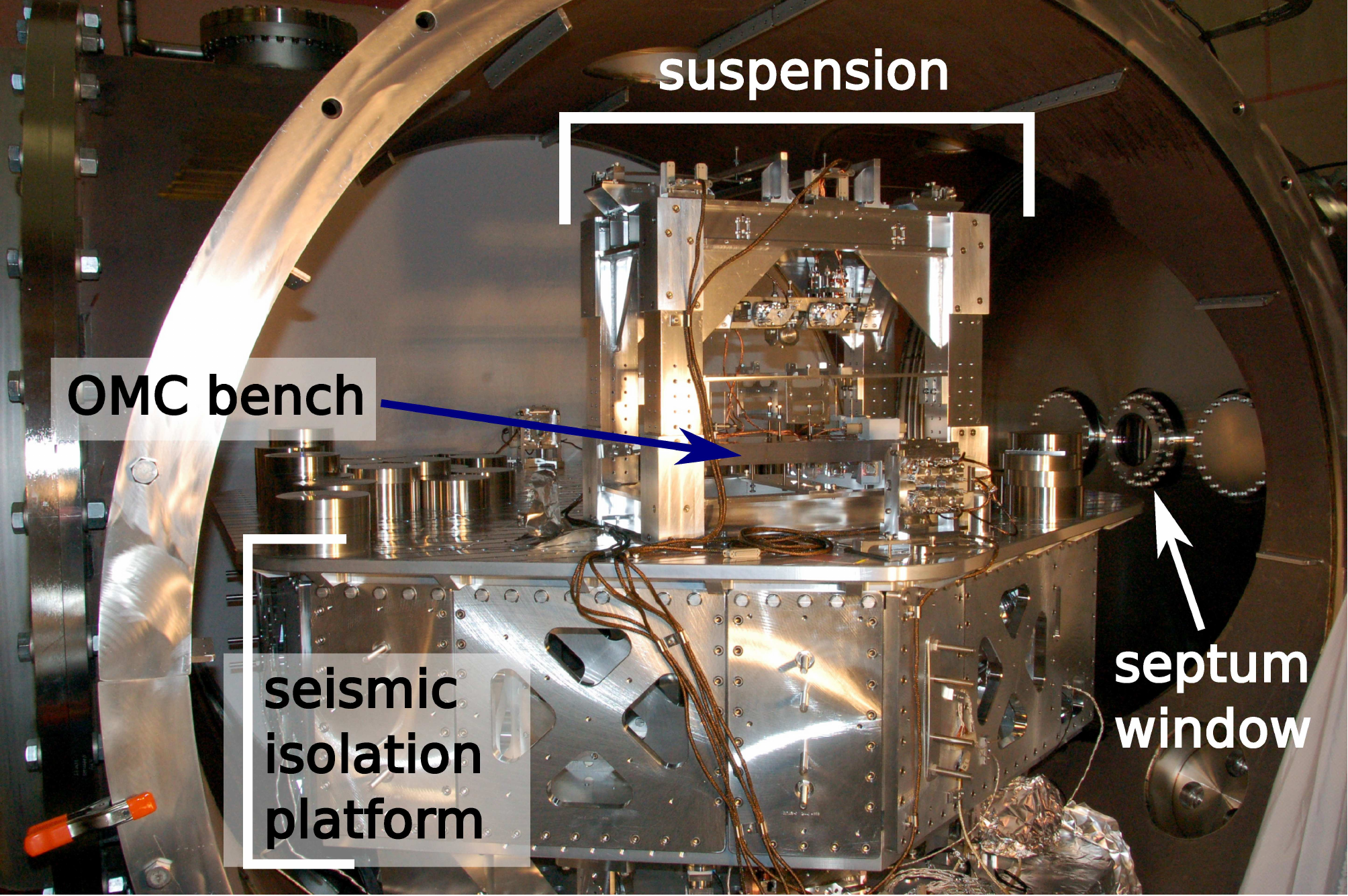}
}
\caption{(a) Diagram illustrating the design of the monolithic OMC
  bench; (b) Photograph of the installed output mode cleaner,
  suspension, and seismic isolation platform. The OMC is located in a
  dedicated vacuum chamber, separated from the main vacuum enclosure
  by a septum window, allowing rapid venting cycles during
  commissioning.}
\end{figure}

\begin{figure}[t]
\centerline{\includegraphics[]{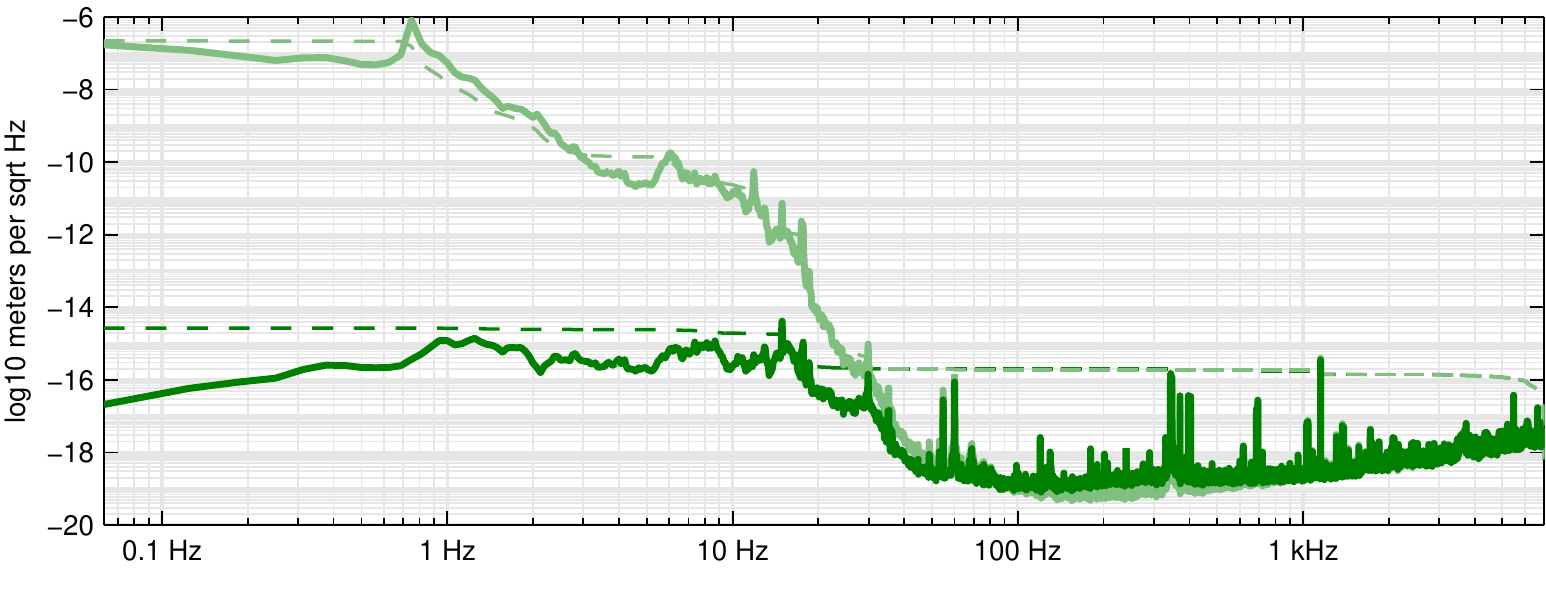}}
\caption{\label{fig:residual-DARM}Calibrated DARM displacement (upper
  traces) and residual motion (lower traces), shown as amplitude
  spectral density (\full, in meters per $\sqrt{\textrm{Hz}}$)
  integrated root-mean-square displacement (\dashed, in meters).  The
  RMS DARM displacement above 0.07 Hz is reduced by approximately eight
  orders of magnitude by the control system in order to remain
  sufficiently near the operating point.  The residual
  motion sets a lower limit for the DARM offset in DC readout to avoid
  fringe-wrapping.}
\end{figure}

The Enhanced LIGO output mode cleaner was installed in a dedicated
vacuum chamber (photograph in figure~\ref{fig:OMC-chamber}) at the
interferometer's output port.  Of the power arriving at the output
port, 97\% is directed to this new DC readout path while the remaining
3\% is directed to the existing RF readout system.  The control system
used to maintain resonance uses the RF readout when initially bringing
the system from an uncontrolled to a controlled (`locked') state, just
as in initial LIGO\cite{Evans2002Lock}.  Once the interferometer is
held on resonance by the control system, we introduce an offset in the
DARM degree of freedom, which produces carrier light at the output
port.  (The possibility of using a MICH offset to produce the local oscillator
was not investigated for Enhanced LIGO, as modeling showed that a
pure DARM offset would be effective.)
The output mode cleaner is brought into resonance with this
carrier light and then sensing of the DARM degree of freedom is
transferred from the RF readout to the photodiodes sensing the light
transmitted through the OMC.

%% DARM offset: how to choose
To produce a sufficiently strong and stable local oscillator, the DARM
offset must significantly exceed the RMS residual arm motion ($\sim
0.1$ picometers, depicted in figure~\ref{fig:residual-DARM}), produce
enough power at the output port such that the contrast defect becomes
negligible, and provide a signal whose shot noise exceeds the
electronics noise of the readout. 
Smaller DARM offsets are preferred, as a larger DARM offset results in
diminished power recycling gain due to the power exiting the dark
port, and because larger DARM offsets generally increase noise
couplings.  An optimum can thus be found.

In the homodyne system, a $\sim30-100$ mW local oscillator at the laser
carrier frequency is instead introduced by offsetting the differential
arm length (DARM) very slightly ($\sim 10$ picometers, equivalent to
60 microradians or 0.4 Hz detuning of each arm cavity, or 0.01 radians
at the beamsplitter) from the dark fringe.  The RF sidebands are no
longer used for length sensing at the output port, but they still
serve to provide sensing of some angular degrees of freedom.

\section{Mode Cleaner Design and Construction}

A four-mirror bow-tie arrangement (depicted in
figure~\ref{fig:omc-diagram}) was chosen for the mode cleaner design.
This non-colinear design prevents direct reflection of rejected light
back into the interferometer.  A design with an even number of mirrors
was preferred
as it has a sparser density of higher-order-mode resonances.
Selecting the angle of incidence of the beam on the cavity mirrors is
a compromise between avoiding small-angle scattering (by using a high
angle of incidence) and avoiding the introduction of too much
astigmatism.

A higher cavity finesse results in better filtering ability, but also
magnifies the effect of any intra-cavity losses in the OMC.  The
cavity design maximizes the finesse subject to the constraint that
transmission of the carrier power should exceed 99\%, assuming
roundtrip cavity power losses of 100 ppm.

The cavity length was chosen to provide adequate attenuation of the RF
sidebands, and its geometry ($g$-factor) was chosen to sufficiently
attenuate higher-order modes.  The optimal $g$-factor depends on the
specific details of the frequency and spatial spectrum of modes at the
output port.  These depend strongly on the details of the
interferometer optics, alignment, and thermal state. 
To deal with these unknown factors, we designed the OMC using a model
in which the power in each higher order mode was proportional to
$1/n^2$, where $n$ is the sum of the horizontal and vertical TEM mode 
indicies, and where the RF sidebands had their nominal power.  The designed
and as-built properties of the output mode cleaner cavities are given
in Table~\ref{tab:OMCproperties}.

One disadvantage of the chosen design is that the 4-th order mode is
nearly degenerate with the fundamental mode.  We did experience
problems with accidental degeneracy in one of the mode cleaners, which
was addressed by changing the operating temperature of the thermal
actuator (which had a small coupling to the effective radius of
curvature of the mirror).  The next version of the output mode cleaner
will be designed with a slightly different g-factor to avoid this
problem.

The cavity was constructed by rigidly mounting the cavity optics
to a baseplate, similar to the LISA optical bench
design\cite{dArcio2010Optical}.  The baseplate is a slab of Corning ULE glass $450 \mathrm{mm}
\times 150 \mathrm{mm} \times \mathrm{39}\mathrm{mm}$; components were
bonded using Optocast {\sffamily 3553LV-UTF-HM} UV-cure epoxy.
Two of the cavity mirrors are outfitted with position actuators: a
fast, short-range ($\lesssim0.1$ \micron) PZT, and a slow, long-range
($\approx 20$ \micron) thermal actuator consisting of a 1 inch segment
of aluminum tube warmed by a resistive heater.

%% Cavity properties [Table]
% These values are all from Sam's document T080144:
% https://dcc.ligo.org/cgi-bin/private/DocDB/ShowDocument?docid=5416
\begin{table}
\centering
\begin{tabular}{l l l l l l}
\hline 
parameter          & symbol & design      & H1          & L1            & units   \\                    
\hline
perimeter          &        & 1.042       & 1.077       & 1.016         & m       \\
beam waist         & $w$    & 477         & 496         & 463           & \micron \\
finesse  & $\mathcal{F}_\textrm{omc}$ & 400         & 360         & 360           &         \\
free spectral 
range              & $\nu_0$     & 287.7       & 278.3       & 295.2         & MHz     \\  % fsr = c/p
cavity pole        &             & 360         & 390         & 410           & kHz     \\  % f_c = fsr/(2F)
$g$-factor         & $g$         & 0.739       & 0.725       & 0.722         &         \\
higher order mode 
spacing            &             & 69.4        & 67.2        & 71.8          & MHz     \\
%transmission       & 1           & $\geq$0.95  & $\geq$0.90    &         \\
\hline
\end{tabular}
\caption{Designed and measured properties of the Hanford and Livingston output mode cleaners.}
\label{tab:OMCproperties}
\end{table}

%% OMC Suspension
To isolate the mode cleaner from environmental disturbances, the
optical bench was hung as the second stage in an actively-damped
double-pendulum suspension system,
%\cite{Robertson2006Conceptual,Robertson2009OMC} non-public
which was in turn mounted on an active isolation
system\cite[Chapter 5]{KisselThesis}.  The active isolator, and the
OMC and its suspension system, were all located in vacuum.

%% OMC Photodiodes
The photodiodes were also mounted on the OMC baseplate, and read out
by in-vacuum preamplifiers.  The output from the mode cleaner was
split via a 50/50 beamsplitter and directed to two Perkin Elmer 3mm
diameter InGaAs photodiodes (part number C30665GH), with measured
quantum efficiency $> 0.95$ at 1064 nm.  The photocurrent was
converted to voltage across $100 \Omega$ transimpedance. Subtraction
of the signals from the two photodiodes produces a diagnostic
``nullstream'' containing the anti-correlated component of the PD
signals, while the sum of the photodiode signals contains the desired
DARM sensing.
The rigid mounting of the PDs to the OMC baseplate reduces the
possibility of beam motion coupling to photocurrent through photodiode
nonuniformities, and the in-vacuum preamplifiers reduce the effect of
electronic or triboelectric noises.

%% Front-end Computers 
The mode cleaner was controlled and the DC readout signals were
acquired using a prototype of the Advanced LIGO real-time digital
signal processing system, operating at 32768 samples per second.
%equipped with analog-to-digital and digital-to-analog converters and
%interfaced to other systems via reflected memory over a fiber ring and
%EPICS over ethernet, operating at 32768 samples per second (Hz)
% \cite{Bork2009ELIGO}, \cite{Bork2009AdvLigo} -- not public
The LIGO Realtime Code Generator\cite{Bork2009ELIGO} allowed fast
prototyping and implementation of complex servos.

%\section{Cavity length control}
% Nobody seems to like one-paragraph sections.

The mode cleaner cavity length was locked by modulating its length at
$\sim10$ kHz using the PZT actuator and synchronously demodulating the
transmitted light signal at the same frequency.  Fast corrections were
applied to the PZT and then gradually offloaded onto the thermal
actuator.

% PF: Give numbers for amplitude of the modulation; what determines it?

\section{Alignment system}

The beam from the interferometer output port is directed to the OMC by
two suspended, steerable mirrors (depicted in
figure~\ref{fig:omc-diagram}), prototypes of the Advanced LIGO
steering mirrors described in~\cite{Slagmolen2011TipTilt}.  These mirrors
allow continuous actuation on the alignment of the beam from the
interferometer to the OMC.

The same spurious higher order modes that motivate the use of an
output mode cleaner also make it challenging to sense the best OMC
alignment.  A servo system that maximizes the intensity of the
transmitted light is not optimal, since the presence of higher order
modes will cause it to displace the input beam to convert some of the
higher order modes into the cavity mode.  Instead, a servo system was
implemented to maximize the transmission of the arm cavity mode
(containing the GW signal) through the OMC.
This servo operates by driving one of the arm cavity end mirrors at
high frequency (9 kHz) while simultaneously modulating the two OMC
beam steering mirrors at low frequency (1-4 Hz) in both pitch and yaw.
The high-frequency modulation is accomplished by exciting the drumhead
mechanical resonance of the cavity end mirror.
The steering mirror modulation produces approximately 3 microns and 5
microradians of beam motion at the OMC cavity waist.  An error signal
is developed by measuring the power in the 9 kHz line and demodulating
this signal at the steering mirror modulation frequencies.  
An alternate alignment scheme, which maximizes the signal to noise
ratio rather than the signal alone, has been proposed in
\cite{Smith2011Optimal}.

% PF suggested deleting mention of the beam diverter:
%% One of the steering mirrors also serves as a fast beam-diverter,
%% diverting the input beam less than 1 millisecond after a lock-loss to
%% avoid exposing the detection photodiodes to the high-power transient, 
%% which would otherwise destroy the diodes.

\section{Results}

\subsection{Sensitivity}
The DC readout is dominated either by quantum shot noise or by signal
(interferometer displacement noise) at all frequencies except
sometimes near a few mechanical resonances causing beam motion (jitter),
which couples linearly or bilinearly to the readout.

\begin{figure}[t]
\includegraphics[]{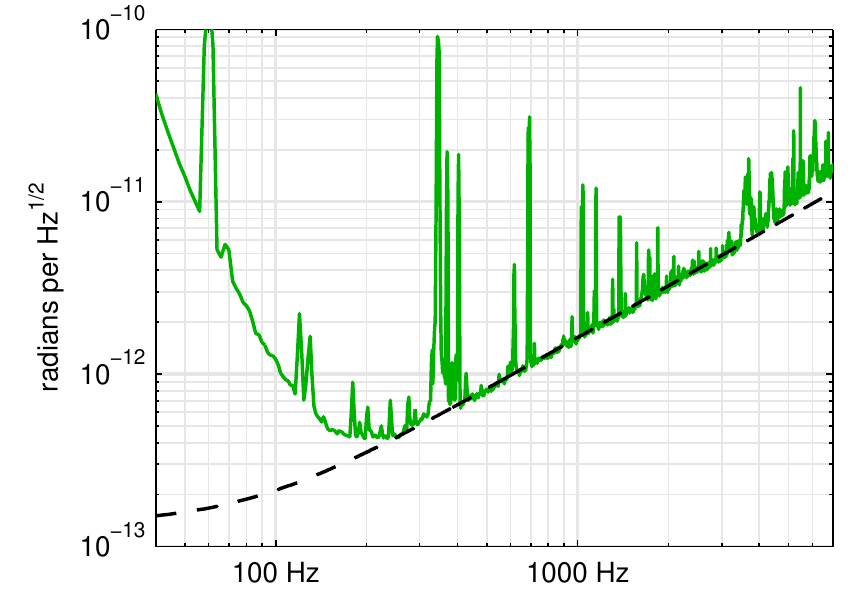} \hfill
\includegraphics[]{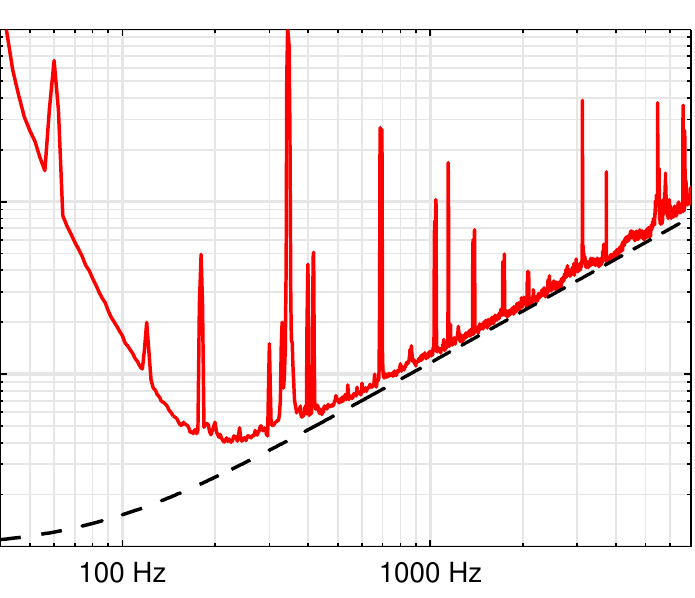}
\caption{\label{fig:shot-noise-comparison}
  Achieved sensitivity (noise floor) amplitude spectral density for
  the Livingston (left) and Hanford (right) detectors (\full), 
  with expected shot noise limit (\dashed).  The Livingston spectrum
  is from 2010-08-11 06:34 UTC and the Hanford spectrum is from
  2010-07-04 08:52 UTC.  The estimated uncertainty in the modeled
  curve is approximately~ $\pm15\%$.
}
\end{figure}

The shot-noise-limited noise floor depends both on the optical gain
and the power at the AS port, making this a good measure of performance
of the system. Both the presence of excess light at the
detection port or sub-optimal optical gain will degrade the
shot-noise-limited sensitivity.   The expected shot noise amplitude spectral density, expressed as a displacement $x_\textrm{shot}$ and an optical phase $\phi_\textrm{shot}$, is 
\begin{eqnarray}
x_{\textrm{shot}}(f)    &= 
\frac{1}{4}\sqrt{\frac{\lambda h c}{2\epsilon P_{IN}}}\ \frac{1}{g_{cr} \mathcal{F_\textrm{arm}}}
\ \left|1 + i \frac{4\mathcal{F_\textrm{arm}}L}{c} f\right|\\
\phi_{\textrm{shot}}(f) &= \left(\frac{2\pi}{\lambda}\right)x_\textrm{shot}(f)
\end{eqnarray}
where $h$ is Planck's constant, $c$ is the speed of light, $P_{IN}$ is
the power delivered by the laser source, ${g_{cr}}^2$ is the power
recycling gain, $\mathcal{F_\textrm{arm}}$ is the finesse of the arm cavities,
$\epsilon$ is the input and output
efficiency of the interferometer, and $\lambda$ is the laser wavelength.
The parameters and efficiency used
in the model are given in table~\ref{tab:ifoparams}.

%% c = 299792458;
%% L = (3995.032 + 3995.001)/2;
%% fsr = c/(2*L);
%% finesse_H1 = fsr / (2 * 83.7)
%% finesse_L1 = fsr / (2 * 85.6)

\begin{table}[t]
\centering
\begin{tabular}{l l l l l}
\hline 
parameter           &  symbol           &  H1          & L1       \\ 
\hline
input power         &  $P_{IN}$         & 20.27 W      & 11.65 W  \\
arm cavity pole     &  $f_c$            &  83.7 Hz     & 85.6  Hz \\
finesse    & $\mathcal{F}_\textrm{arm}$ &  224         & 219      \\
power recycling gain&  ${g_{cr}}^2$     &  59          & 41       \\
\hline
carrier fraction after
phase modulation    &  $J_0(\Gamma)^2$  & 0.94         & 0.95 \\
input optics        &                   & 0.82         & 0.75  \\
interferometer mode-matching &          & 0.92         & 0.92  \\
output faraday isolator transmission &  & 0.94         & 0.98  \\
DC readout pickoff fraction  &          & 0.953        & 0.972 \\
OMC mode-matching            &          & 0.70         & 0.95  \\
OMC transmission and 
PD quantum efficiency        &          & 0.95         & 0.95  \\
\hline
net power efficiency         &$\epsilon$& 0.42         & 0.56  \\
\hline
\end{tabular}
\caption{\label{tab:ifoparams}Interferometer parameters used in the
  shot noise model.}
\end{table}

The achieved sensitivity is compared to the expected shot noise limit
in figure~\ref{fig:shot-noise-comparison}; we find that the observed
shot noise limit is consistent with the model.  The agreement
indicates that there is no significant excess shot noise due to
higher-order modes, and the optical gain is as predicted; the benefits
of homodyne detection with an OMC are achieved.  However, sub-optimal
transmission through the OMC due to imperfect
mode-matching 
of the interferometer spatial mode into the OMC 
and/or increased OMC intra-cavity losses
was one
of the largest inefficiencies in the Hanford detector, equivalent to a
30\% power loss and commensurate 15\% increase in shot noise,
partially negating the effect of Hanford's higher input power (see
table~\ref{tab:ifoparams}).

\subsection{Noise couplings}

Noise couplings from the laser and RF oscillator are modified
substantially by DC
readout\cite{Ward2008DC,RobWardThesis,Somiya2006Frequency}.  Other
noise couplings (of displacement noises, in particular) generally
remain the same.  These couplings were measured at both Hanford and
Livingston and compared to models.

To model the laser and oscillator noise transfer functions, we used
the plane-wave frequency-domain interferometer simulation tool
Optickle\cite{Evans2007Optickle}.  The simulation includes realistic
parameters such as contrast defect and arm cavity finesse imbalance,
and also includes the cross-couplings due to the servo control
systems.  However, the model does not include effects due to
mode-matching or higher-order spatial modes.

DC readout also introduces at least two new noise sources: OMC path
length fluctuations and beam pointing noise (jitter).

\subsubsection{OMC path length noise}
%% Length noise
% Good elogs on the subject:
%  'OMC length noise coupling'              - L1 2009-01-08 - http://goo.gl/H4f6m
%  'Calibrated OMC_LSC_I_OUT to femtometer' - L1 2009-03-09 - http://goo.gl/c4ylN

Near resonance, the transmission of the OMC is a quadratic function of
the intra-cavity optical path length.  A change in OMC transmission is
indistinguishable from a DARM perturbation; thus OMC path length noise
can appear in the readout signal.  Because the coupling is quadratic
the contribution to the readout depends on the particular spectrum of
path length noise.  In practice, the dominant contribution tends to be
via bilinear upconversion of low-frequency motion.  The
root-mean-square deviation from resonance was found to be
$\sim75\times10^{-15}$~m.  

By adding band-limited random noise to the OMC PZT actuator and
observing the effect in the transmitted light, and by comparing this
coupling to the typical spectrum of the OMC path length error signal,
we we estimated that the OMC path length noise contribution to DARM
was further than $10\times$ below the noise floor.

% Should add some words about why cavity length noise is less of a
% problem than cavity alignment noise.  Need to design the cavity so
% that other modes are far away from the desired resonant mode.

%% \begin{figure}[b]
%% \centerline{\includegraphics[width=\columnwidth]{figures/waldman-1236648503.pdf}}
%% \caption{\label{fig:length-noise} Cavity length noise.  To-do: Maybe the ``noise-budget'' style plot would be better?}
%% \end{figure}

\subsubsection{Beam jitter}
Beam jitter is perhaps the most important new noise source introduced
in the DC readout system.  The OMC converts motion of the incident
beam into amplitude modulation, which pollutes the readout.  In the
ideal case, this would be a quadratic effect, but the presence of
spurious higher order modes introduces both linear and bilinear
contributions.
We have several means of controlling the beam jitter contribution to the
readout.  In practice, we utilize all of them:

\begin{itemize}
\item \emph{Remove mechanical resonances and increase isolation.}  Several
  prominent beam-jitter peaks in the readout spectrum were removed by
  replacing a fixed steering mirror with a suspended one; and by
  adding additional vertical isolation to the steering mirrors.

\item \emph{Cancel the motion.}  Motion at the 60 Hz power-line
  frequency is introduced via magnetic coupling to the magnets used to
  actuate on the suspended steering mirrors.  We implemented a feed-forward
  correction using signals derived from a magnetometer located just outside
  the vacuum chamber.

\item \emph{Place steering optics where their coupling to beam jitter
  is small.}
  Beam propagation geometry can magnify the effects of optic motion to 
  beam motion.
  For a given amount of optic jitter, the coupling to readout noise can
  be reduced by proper design of the mode-matching telescope.
  This constraint is at odds with the desire for maximum controllability
  of the input beam pointing.  
  In practice the requirement is that sufficient actuation range be 
  provided, and that the optics be well separated in Gouy phase to
  create a non-degenerate control matrix.

\item \emph{Reshape the beam.}  Linear beam jitter coupling arises due
  to the presence of spurious carrier light in the Hermite-Gauss TEM01 
  mode incident on the mode cleaner; displacement of the beam couples
  this mode into the fundamental Gaussian TEM00 mode of the mode cleaner
  cavity.
  By introducing offsets into the interferometer's global angular
  sensing and control (ASC) system\cite{Barsotti2010Alignment}, this
  spurious TEM01 mode can be minimized, reducing the linear
  sensitivity to beam motion.  During Enhanced LIGO this was found to
  be a highly effective technique, but it was not automated.
\end{itemize}

\begin{figure}[t]
\includegraphics[width=0.33333\columnwidth]{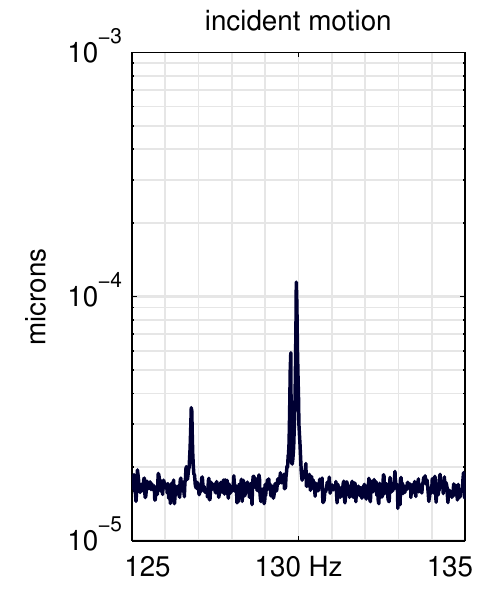}
\includegraphics[width=0.33333\columnwidth]{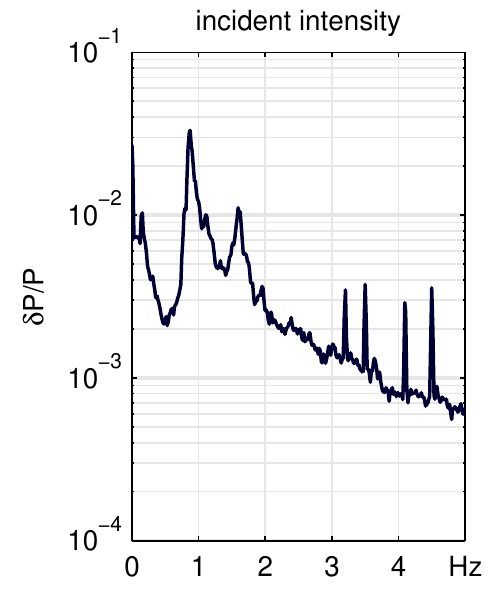}
\includegraphics[width=0.33333\columnwidth]{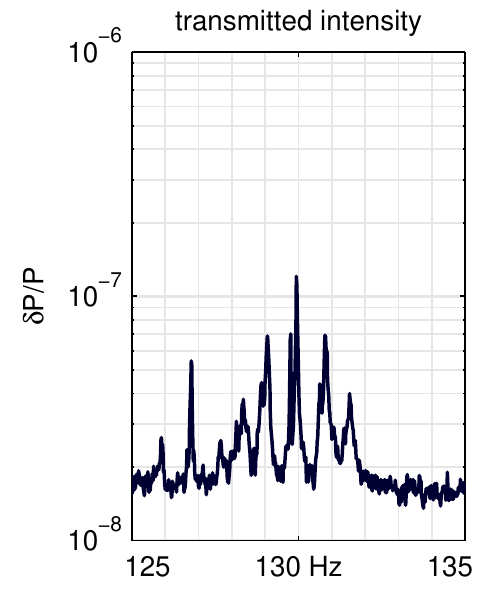}
\caption{\label{fig:jitter-mechanism}Example of linear and bilinear
  beam jitter coupling in the L1 OMC.  Left: the light incident on the
  OMC shows motion at $\sim 130$ Hz.  Center: the incident light
  contains intensity modulations at $\sim 0.87$ and $\sim 1.6$
  Hz. Right: the light transmitted through the OMC contains intensity
  modulations at $130$ Hz, $130 \pm 0.87$ Hz, and $ 130 \pm 1.6$ Hz.
  (The sources of the incident beam motion and intensity fluctuations
  at these frequencies have not been identified.)}
\end{figure}

An example of the observed beam jitter coupling is shown in
figure~\ref{fig:jitter-mechanism}.  A prominent spectral line at
$\sim130$ Hz appears in quadrant photodiode (QPD) signal, indicating
relative motion in yaw between the incident beam and the OMC.  This
line also appears in the signal of the light transmitted through the
OMC, indicating some linear coupling of beam jitter to transmission.
The transmitted spectrum also contains sidebands around the 130 Hz
line at separations of $\pm0.875$ Hz and $\pm1.6$ Hz which arise due
to bilinear coupling.

\subsubsection{Oscillator noises}

\begin{figure*}
% Oscillator AM
\subfloat[Oscillator amplitude noise coupling (L1)]{
  \includegraphics[]{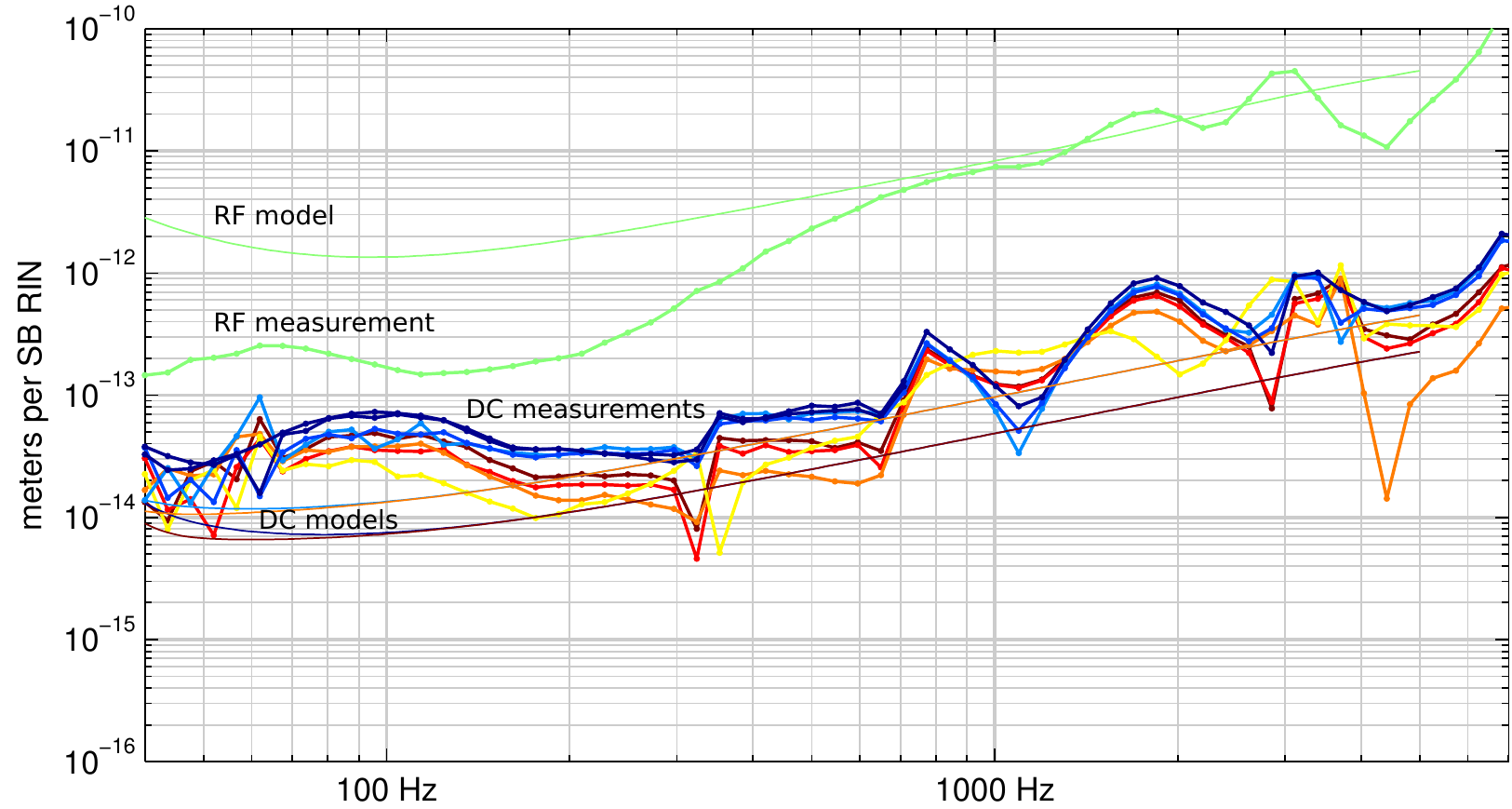}
  \label{fig:osc-AM-L1}  
}

\subfloat[Oscillator amplitude noise coupling (H1)]{
  \includegraphics[]{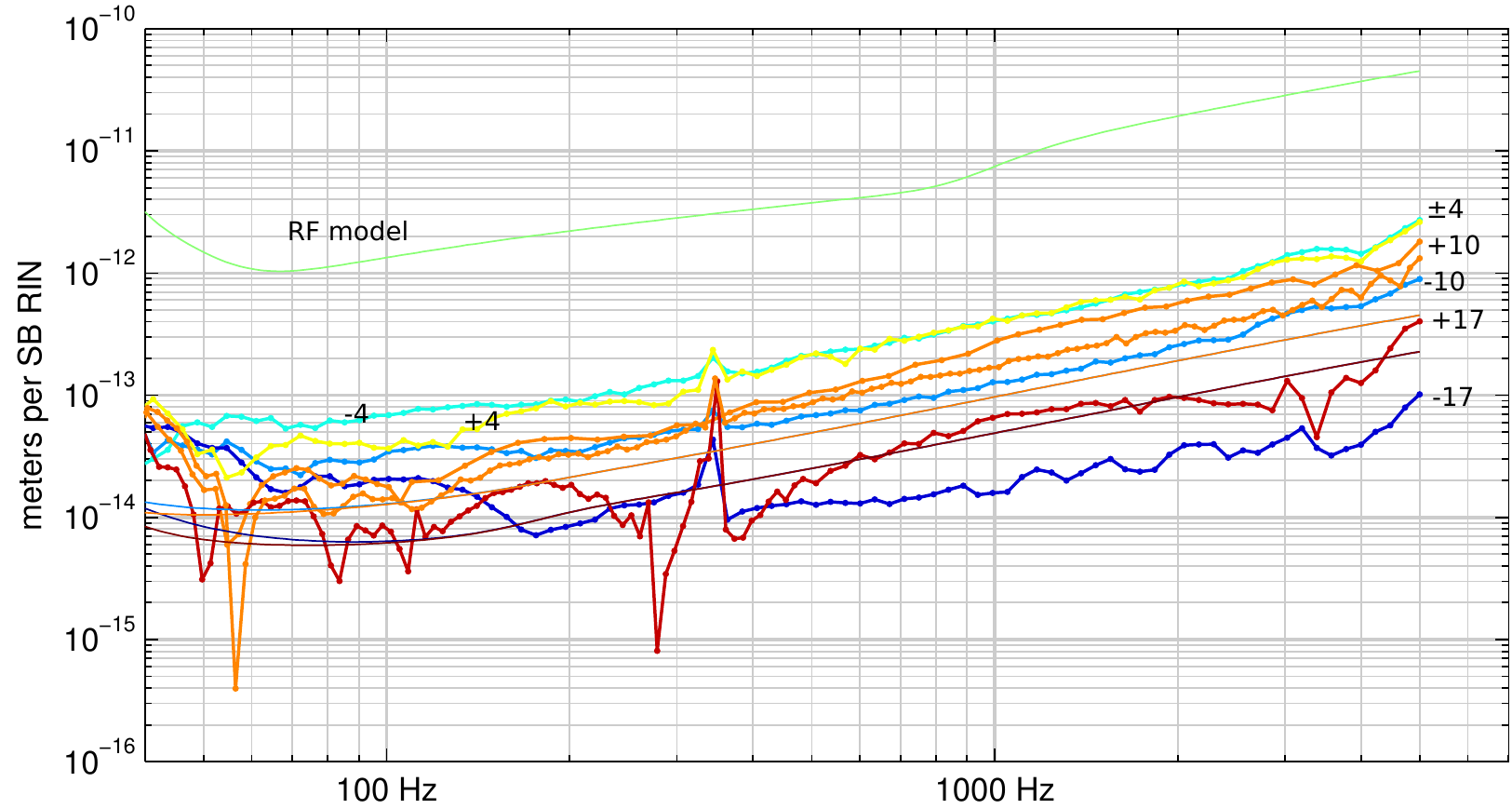}
  \label{fig:osc-AM-H1}
}

\caption{\label{fig:osc-AM}Oscillator amplitude noise couplings
  to the gravitational wave readout channel.  Solid lines are the
  results of a frequency-domain, plane-wave model; dotted lines are
  linear transfer function measurements.  Measurements were taken
  with DARM offsets between $-20$ and $+20$~pm, and with RF readout (L1 only).
  Color represents the DARM
  offset, with warm colors for positive offsets and cool colors for
  negative offsets.  Measurements and models for RF readout are in
  green.
  The vertical axis is equivalent DARM meters per relative intensity noise ($\delta P/P$)
  of the RF sideband.
}
\end{figure*}

\begin{figure*}
% Oscillator PM
\subfloat[Oscillator phase noise coupling (L1)][Oscillator phase noise coupling (L1)]{
  \includegraphics[]{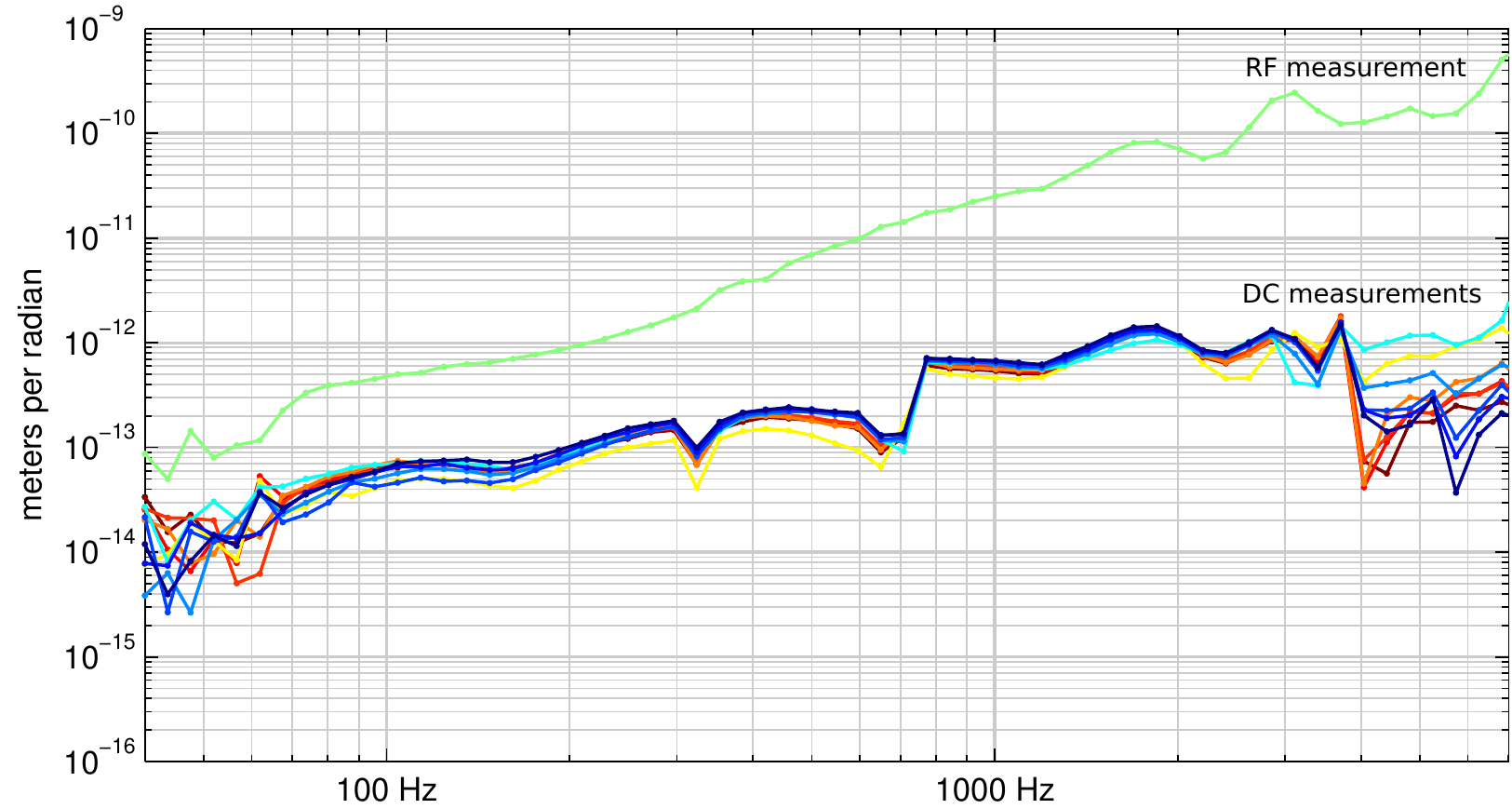}
  \label{fig:osc-PM-L1}
}

\subfloat[Oscillator phase noise coupling (H1)][Oscillator phase noise coupling (H1)]{
  \includegraphics[]{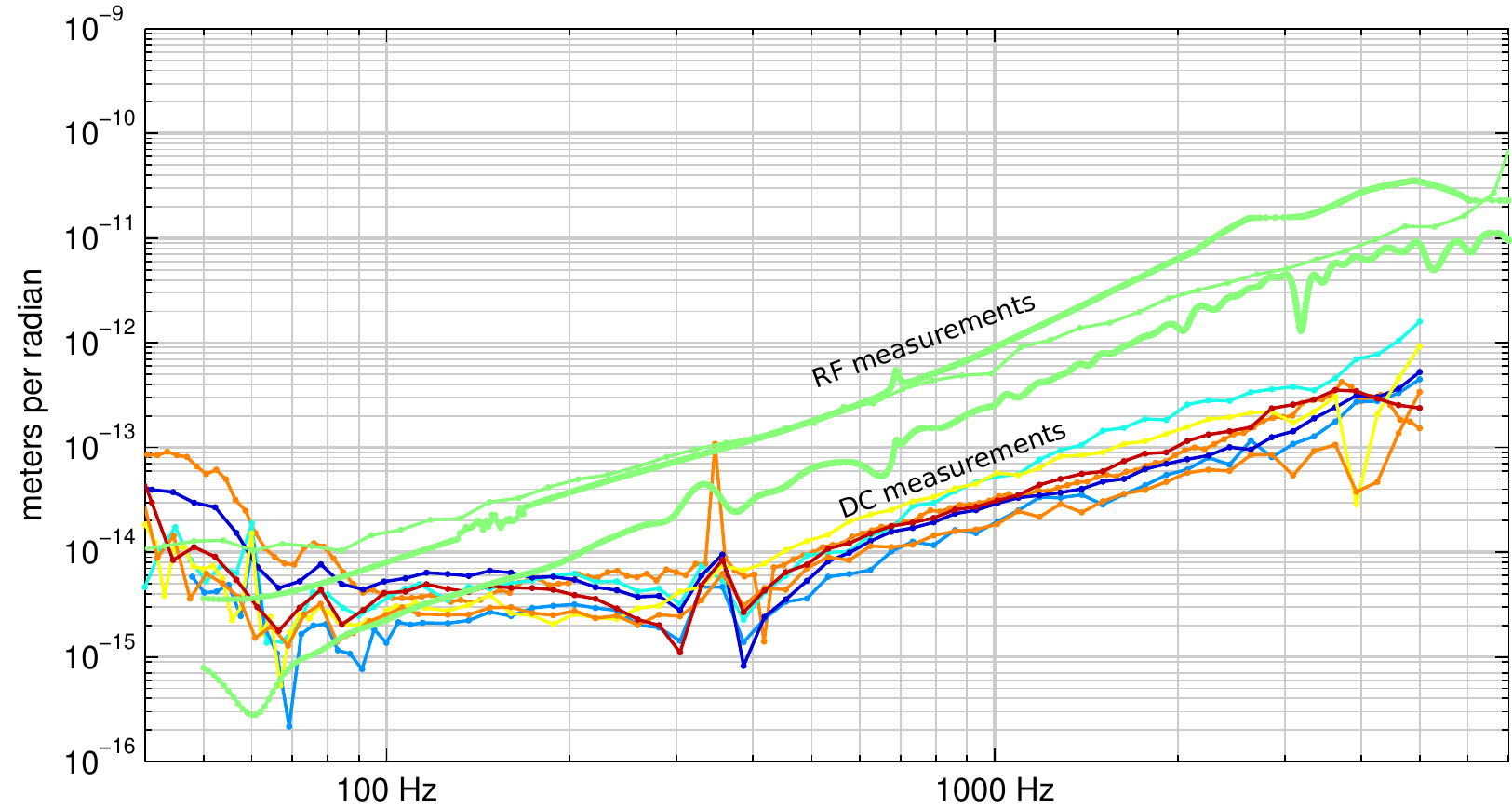}
  \label{fig:osc-PM-H1}
}

\caption{\label{fig:osc-PM}Oscillator phase noise couplings to the
  gravitational wave readout channel, calibrated in equivalent
  displacement (meters).  All traces are linear transfer function
  measurements.   Measurements were taken
  with DARM offsets between $-20$ and $+20$~pm, and with RF readout. Color represents the DARM offset, with warm colors
  for positive offsets and cool colors for negative offsets.
  Measurements and models for RF readout are in green.}
\end{figure*}

Although DC readout does not depend directly on the RF oscillator, the
RF sidebands are still present in the interferometer (in particular at
the output port, before the OMC) and are used to sense other degrees
of freedom.  Noise on the RF sidebands can couple to DC readout both
directly (by off-resonance transmission through the OMC) and
indirectly (via cross-couplings from other loops).

Oscillator amplitude modulation (AM) creates a time-varying modulation depth in the phase
modulator used to produce the RF sidebands.  In the frequency domain, this appears as anti-correlated
AM on the carrier and the RF sidebands at the input to the
interferometer.  These sidebands around the carrier and around the RF
sidebands propagate differently through the interferometer and through
the OMC: the sidebands around the carrier are filtered by the coupled
cavity pole while those around the RF sidebands are not; conversely,
the sidebands around the RF sidebands are strongly attenuated by the
OMC while those around the carrier are transmitted.  Varying carrier
intensity also leads to a displacement noise via radiation pressure
acting on the differentially detuned arm cavities.
Oscillator phase noise, on the other hand, creates phase modulation
sidebands around the RF sidebands, but does not produce any modulation
of the carrier.

%The electro-optic modulators also produce a small amount of AM
%themselves; the measured ratio of AM to PM modulation depths of the
%phase modulator is approximately $10^{-4}$.

To measure the couplings of oscillator noises, we temporarily reverted
to using a general purpose (IFR 2023A) RF function generator instead
of the crystal oscillator as the interferometer's 25 MHz oscillator.
This function generator was configured to impress amplitude and phase
modulation during swept-since measurements of the linear transfer
functions from amplitude/phase modulation to the DC readout signal.

The results are depicted in figures~\ref{fig:osc-AM} and
\ref{fig:osc-PM}. We find that the DC readout sensitivity to RF
oscillator amplitude and phase noises are reduced by a factor of
10-100 as compared to RF readout.  The oscillator amplitude modulation (AM) coupling agrees
well with the model.  (We do not attempt to model the oscillator phase
noise coupling, since it is strongly influenced by effects not modeled
by Optickle.)

\subsubsection{Laser noises}

\begin{figure*}[]
% Laser AM
\subfloat[Laser amplitude noise coupling (L1)][Laser amplitude noise coupling (L1)]{
  \includegraphics[]{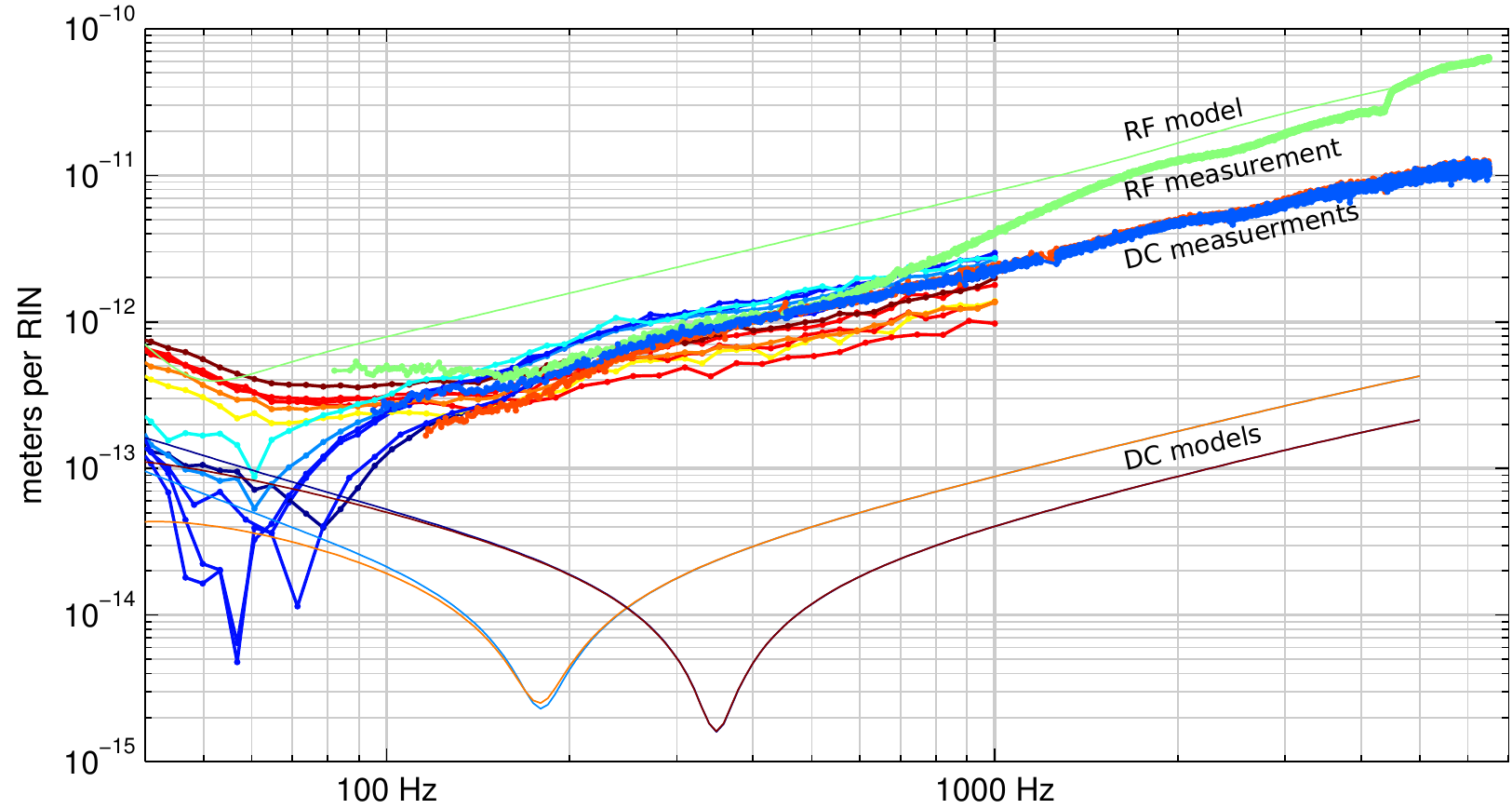}
  \label{fig:laser-AM-L1}
}

\subfloat[Laser amplitude noise coupling (H1)][Laser amplitude noise coupling (H1)]{
  \includegraphics[]{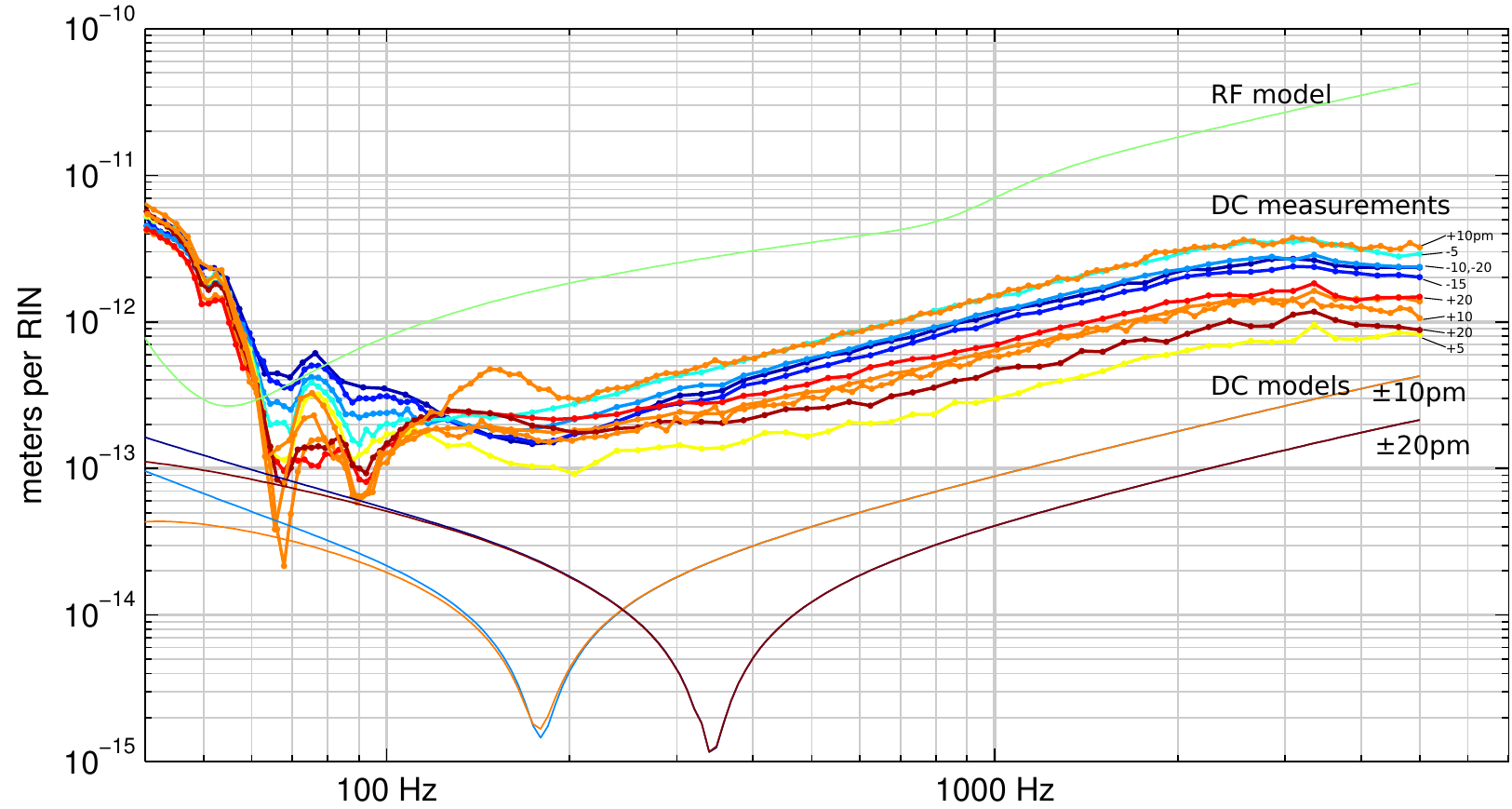}
  \label{fig:laser-AM-H1}
}

\caption{\label{fig:laser-AM}Laser amplitude noise couplings to the gravitational wave readout channel.  Solid lines are the results of a frequency-domain, plane-wave model; dotted lines are linear transfer function measurements.  Measurements were taken at DARM offsets between $-20$ to $+20$ pm and with RF readout (L1 only). Color represents the DARM offset, with warm colors for positive offsets and cool colors for negative offsets.  Measurements and models for RF readout are in green.}
\end{figure*}

\begin{figure*}
% Laser FM
\subfloat[Laser frequency noise coupling (L1)][Laser frequency noise coupling (L1)]{
  \includegraphics[]{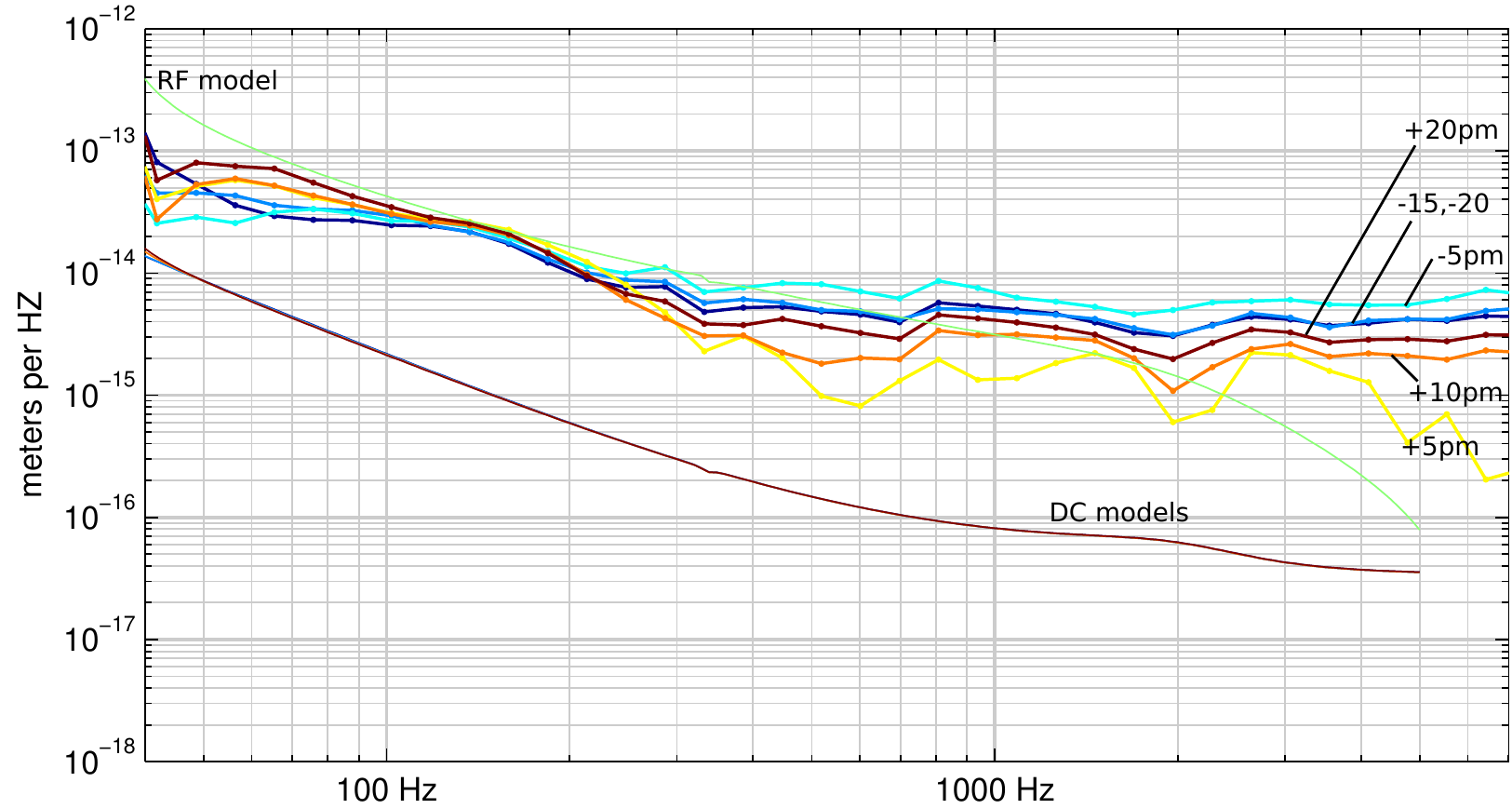}
  \label{fig:laser-FM-L1}
}

\subfloat[Laser frequency noise coupling (H1)][Laser frequency noise coupling (H1)]{
  \includegraphics[]{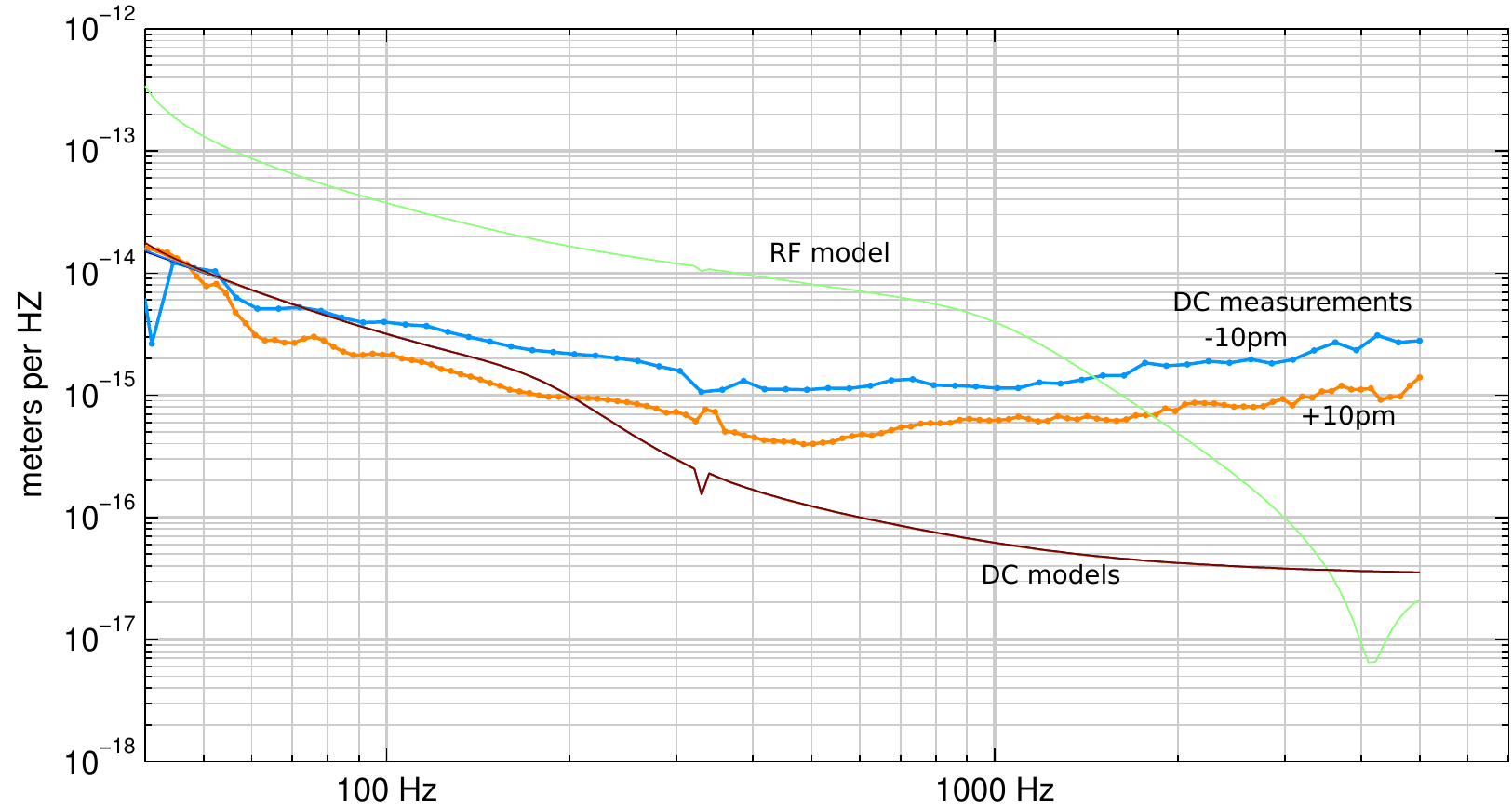}
  \label{fig:laser-FM-H1}
}

\caption{\label{fig:laser-FM}Laser frequency noise couplings to the gravitational wave readout channel.  Solid lines are the results of a frequency-domain, plane-wave model; dotted lines are linear transfer function measurements.  Measurements were taken with DARM offsets of $\pm20, \pm10,$ $\pm5$~pm at L1 and $\pm10$~pm at H1. Color represents the DARM offset, with warm colors for positive offsets and cool colors for negative offsets.}
\end{figure*}

Noise on the laser source appears on both the carrier and the RF
sidebands. As in the discussion of oscillator noises, modulation of
the carrier is attenuated by the coupled cavity pole while noise
around the RF sidebands is attenuated by the OMC.  Amplitude noise
couples directly to DC readout, while frequency noise must be
converted to AM by asymmetries in the interferometer to couple to the
readout.

To measure the coupling of laser intensity noise, the intensity of the
laser source was modulated by driving the error point of the laser
intensity stabilization servo.
The coupling of laser frequency noise to the readout was measured by
driving the error point of the common mode servo (CARM), which nulls
the mismatch between the laser frequency and the average of the two
arm lengths by actuating on the laser frequency.  We calibrate the
CARM sensor by injecting a sinusoidal excitation into the drive to one
of the end mirrors.  By measuring the response in both CARM and DARM
and compensating for the suppression due to the feedback control
system, the DARM calibration is ported to CARM, which is then be
converted from meters to Hz using the relation $\delta{}L/L =
\delta\nu/\nu$ with $L= 3995$~m and $\nu=c/\lambda$.

The measured intensity noise and frequency noise transfer functions
are depicted in figures~\ref{fig:laser-AM} and \ref{fig:laser-FM},
respectively.  In both cases, the observed coupling is much greater
than the model predicts.  This coupling may be due in part to residual
transmission of higher order spatial modes
from imperfect alignment of the interferometer output beam to the OMC;
carrier light in these higher order spatial modes does not
resonate in the arms and thus is not attenuated by the coupled-cavity
pole.   Further evidence that this coupling is dominated by
contributions due to higher-order modes is that it changed
significantly between two measurements made a few months apart. 
The
higher-order mode content is highly dependent on the thermal state of
the interferometer and its coupling to DC readout depends on the
alignment of the output beam into the OMC. 

\section{Conclusion}

The Enhanced LIGO experiment has validated the DC readout scheme as a
low-noise readout system scalable to high-power operation and
alleviating some of the troubles experienced with RF readout.  We have
shown that the shot-noise-limited sensitivity achieved with DC readout
is consistent with expectations, and presented measurements of laser
and oscillator noise couplings to the gravitational wave channel.
These couplings are generally much better with DC readout than with RF
readout.  We find that a simple plane wave model is not adequate to
explain the laser noise couplings; future work should utilize a more
complex model incorporating the effects of higher order spatial modes.

\ack

TF thanks the National Science Foundation for support under grant
PHY-0905184 and Caltech for support via the LIGO Visitors Program.
The authors would like to thank Koji Arai, Hartmut Grote, 
David Shoemaker, and Zach Korth for comments on this manuscript; and everyone who 
contributed to the Enhanced LIGO effort.
LIGO was constructed by the California Institute of Technology and
Massachusetts Institute of Technology with funding from the National
Science Foundation and operates under cooperative agreement
PHY-0757058. This article has LIGO Document Number
\href{https://dcc.ligo.org/cgi-bin/private/DocDB/ShowDocument?docid=8442}{P1000009}.

\section*{References}

\bibliographystyle{unsrturletal}
\bibliography{references}

\end{document}